# Robust quantum engineering of current flow in carbon nanostructures at room temperature

Gaetano Calogero[1†*], Isaac Alcón[2†*], Onurcan Kaya[2,3,4], Nick Papior[5], Aron W. Cummings[2*], Mads Brandbyge,[6] and Stephan Roche[2,7]

[1]CNR Institute for Microelectronics and Microsystems (CNR-IMM), Zona Industriale, Strada VIII, 5, 95121 Catania, Italy

[2]Catalan Institute of Nanoscience and Nanotechnology (ICN2), CSIC and BIST, Campus UAB, Bellaterra, 08193 Barcelona, Spain

[3]School of Engineering, RMIT University, Melbourne, Victoria, 3001, Australia

[4]Department of Electronic Engineering, Universitat Autonoma de Barcelona (UAB), Campus UAB, Bellaterra, 08193 Barcelona, Spain

[5]Computing Center, Technical University of Denmark, DK-2800 Kongens Lyngby, Denmark

[6]Department of Physics, Technical University of Denmark, DK-2800 Kongens Lyngby, Denmark

[7]ICREA, Institució Catalana de Recerca i Estudis Avançats, 08070 Barcelona, Spain

[†]*equally contributed*

* Corresponding authors: gaetano.calogero@cnr.it (Gaetano Calogero), isaac.alcon@icn2.cat (Isaac Alcón), aron.cummings@icn2.cat (Aron W. Cummings)

**Abstract**

Bottom-up on-surface synthesis enables the fabrication of carbon nanostructures with atomic precision. Good examples are graphene nanoribbons (GNRs), 1D conjugated polymers, and nanoporous graphenes (NPGs), which are gathering increasing attention for future carbon nanoelectronics. A key step is the ability to manipulate current flow within these nanomaterials. Destructive quantum interference (QI), long studied in the field of single-molecule electronics, has been proposed as the most effective way to achieve such control with molecular-scale precision. However, for practical applications, it is essential that such QI-engineering remains effective near or above room temperature. To assess this important point, here we combine large-scale molecular dynamics simulations and quantum transport calculations and focus our study on NPGs formed as arrays of laterally bonded GNRs. By considering various NPGs with different inter-GNR chemical connections we disentangle the different factors determining electronic transport in these carbon nanomaterials at 300 K. Our findings unequivocally demonstrate that QI survives at room temperature, with thermal vibrations weakly restricting current flow along GNRs while completely blocking transport across GNRs. Our results thus pave the way towards the future realization of QI-engineered carbon nanocircuitry operating at room temperature, which is a fundamental step towards carbon-based nanoelectronics and quantum technologies.

**1. Introduction**

The field of single-molecule electronics, which emerged in the 1970s,[1] aims at using single molecules as electronic components within what could be considered the ultimate level of miniaturization in electronics.[2–4] Multiple physico-chemical phenomena have been proposed as a means to modulate current flow through single-molecule devices, such as conformational changes, dipole orientation, spin and charge states, or the formation/breaking of chemical bonds.[5–7] Among these, destructive quantum interference (QI) represents one of the most widely studied mechanisms to tune current at the molecular scale,[8–12] and its archetypical example is a single phenyl ring electrically contacted either in para- or meta-configuration. Within para-contacted rings incoming electrons interfere constructively, enhancing electronic transmission. On the contrary, contacting the two electrodes in meta-position with respect to each other generates destructive quantum interference (QI) within the phenyl unit.[8] Since QI emerges from the wave nature of electrons, the resulting suppression of transmission becomes energy dependent.

In spite of the massive amount of research that has been devoted to advance the field of single-molecule electronics, to date no commercially available technology is yet working at the single-molecule level. This is partly due to the difficulty to electrically contact individual molecules in a robust and reproducible manner which, in turn, arises from the hybrid nature of the interface between the molecule and the metallic bulk electrodes.[2,13]

An alternative approach to exploit the vast physico-chemical versatility of organic molecules for nanoelectronics is to covalently embed them within conductive π-conjugated carbon nanostructures.[14,15] In recent years it has been shown that such carbon nanomaterials may be fabricated with atomic precision via bottom-up on-surface synthesis.[16,17] This approach uses specifically designed organic molecules deposited on metallic surfaces, undergoing various reaction steps at specific temperatures, ultimately leading to the pursued atomically-precise nanostructure.[18] Such an approach permits in a natural way the covalent integration of versatile molecular functionality. For example, bottom-up synthesized graphene nanoribbons (GNRs),[17,19] which are receiving significant attention for nanoelectronics and quantum technologies,[20,21] have been used to electrically probe single molecules via highly robust covalent contacts.[15,22,23] Embedding molecular units hosting QI effects is particularly promising for nanoelectronics, as ultimately it could be used to manipulate current flow with atomic precision. In this regard, the so-called nanoporous graphenes (NPGs), synthesized as a 2D array of laterally connected GNRs, are particularly appealing.[24] NPGs may be thought of as a 2D covalent framework of parallel 1D nanowires (i.e. GNRs) whose electronic coupling depends on their chemical bonding. Quantum transport simulations showed that C-C bonded GNRs, as in the original NPG,[24] are strongly coupled and so injected currents significantly spread through a number of GNR channels.[25] Soon after, it was theoretically shown that if GNRs are bridged via meta-configured phenyl rings, hosting QI, they effectively become electronically decoupled, such that injected currents remain fully confined within the contacted (0.7 nm wide) GNR.[26] Very recently, para- and meta-connected GNRs have been achieved in phenylated NPGs,[27] which highlights the experimental feasibility of this novel approach.

While these initial studies are very promising, the technological use of QI-engineering in future carbon nanoelectronics will require that such effects survive at finite temperature, ideally near or above room temperature. However, to date, all studies evaluating quantum transport in QI-engineered NPGs have focused on the idealized flat (i.e. at 0 K) NPG structures, thus completely neglecting any temperature effects.[26–28] In that regard, though thermal vibrations are known to affect conductance through single-molecule devices,[29,30] different studies have suggested a resilience of QI-induced conductance suppression in the presence of thermal vibrations, both in single molecules[30–32] and graphene nanoflakes.[33] However, other theoretical simulations point in the opposite direction.[34] It is therefore debated which scenario applies to large-scale NPG devices, where an array of QI-modulated molecular units should act collectively while being subject to randomly distributed structural fluctuations at different scales. If QI were functional under these harsh conditions, one could QI-engineer carbon nanocircuitry for multiple applications in micro/nanoelectronics.

In this work, we use numerical simulations to evaluate whether QI-engineering of current flow in NPGs survives under thermal fluctuations at 300 K. To do so, we couple large-scale molecular dynamics (MD) simulations with quantum transport calculations based on the Green's function (GF) formalism. Tight-binding (TB) models equipped with bond-length dependent parameters are used to capture electron-phonon coupling. This methodology, known as the MD-Landauer approach,[35] allows us to calculate the length-dependent electronic transmission, from which we may extract the direction dependent conductivity ($\sigma^\perp$, $\sigma^\parallel$), where $\sigma^\parallel$ is the conductivity in the direction parallel to the GNRs in NPGs and $\sigma^\perp$ is the conductivity in the perpendicular direction. We focus our study on the original NPG,[24] and on the so-called para- and meta-NPGs displaying inter-ribbon connections based on single phenyl rings in either para- or meta-configuration, respectively (Figure 1a).[26] We also include graphene as a reference material with purely 2D isotropic charge transport. This series of materials allows us to disentangle the effect of the various structural features that contribute to transport anisotropy in NPGs. We find that meta-NPG displays the most effective suppression of electronic transport across GNRs, which confirms that the QI embedded in this nanomaterial is fully operational under vibrational disorder. Our results thus highlight the power of QI-engineering to control current flow in π-conjugated carbon nanostructures at room temperature, which is a fundamental requirement for its future technological exploitation.

## 2. Methodology

Large-scale optimizations and MD runs were done using the Airebo force field[36] and the LAMMPS code.[37] The construction of the device geometries and the TB Hamiltonians, starting from the structures generated with MD, was carried out by developing *ad-hoc* Python scripts based on the open-source SISL package.[38] In Section S4 of the SI we describe in detail the required steps to generate the thermally activated structures with MD and to build the device geometries used in the transport simulations. The TBtrans code[39] is used to simulate the transmission spectrum $T_{300K}(E, L)$ or $T_{0K}(E, L)$. Periodic boundary conditions are modelled along the transverse direction x (y) by using 3 (5) k-points in the case of NPG devices or 10 k-points in the case of graphene devices. An energy range $E - E_F$ = [-0.3, 0.15] eV is fixed for all materials, with a resolution of 1.5 meV. Further computational details can be found in Section S4 of the SI. The density functional theory calculations of the band structure for all materials and their response to increasing bi-axial in-plane strains (from 1% to 5%) reported in Figure S1-4 are carried out using the SIESTA code.[40]

## 3. Results and Discussion

### 3.1 Theoretical framework

In order to evaluate the direction-dependent transport properties of each system at 300 K we utilize the MD-Landauer approach.[35] First, we construct 100x100 nm$^2$ periodic samples of each NPG by repeating the primitive cell along both in-plane directions ($x$ and $y$). The resulting large-scale samples are composed of approximately 450,000 atoms depending on the material (see Table S1 in the Supporting Information, SI). We thermalize each sample at 300 K via MD simulations using the Airebo force field[36] as implemented in the LAMMPS code[37] (see Methods for more details) and we extract ten snapshots during the MD run for each material. Each of these snapshots is then smoothly joined to semi-infinite electrodes made of the same material in its pristine form (i.e. at 0 K), as schematically depicted in Figure 1b (see Methods). We automate this procedure by using a series of libraries based on the SISL Python utility,[38] which are available as open access supporting material. For each snapshot, this device integration procedure is repeated for various device lengths ($L$ in Figure 1b) and along each in-plane direction ($x$ and $y$), as required to compute the direction-dependent conductivities (see below).

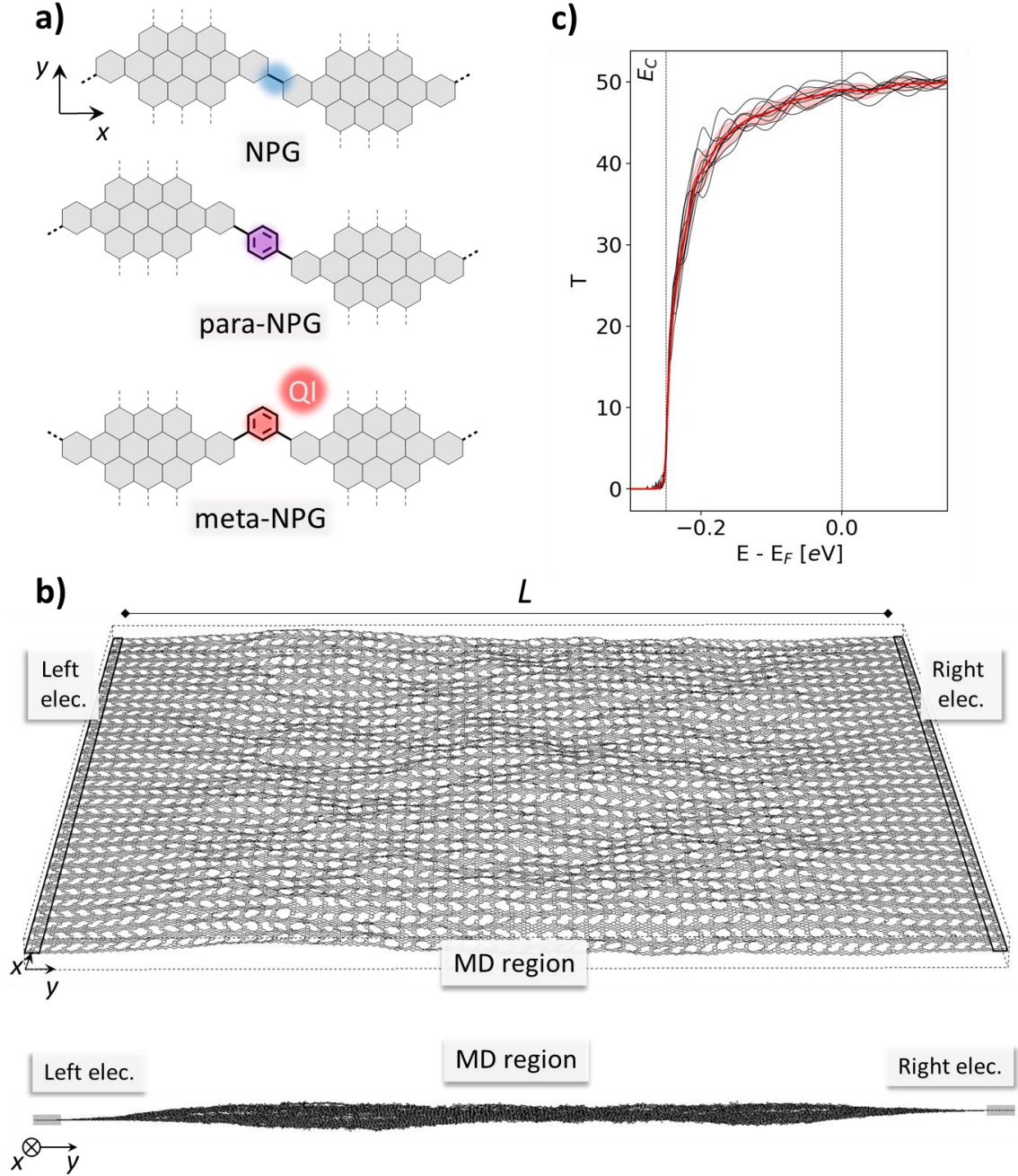

Figure 1. a) Atomic structure of the three considered NPGs characterized by their specific inter-ribbon connection: direct C-C bonding for NPG, para-configured phenyl ring for para-NPG and meta-configured phenyl ring for meta-NPG. b) Schematic of the device geometry after attaching the thermally activated region (300 K) to the pristine electrodes (0 K). As shown in the bottom panel (x-view), a gradually smoothed region connects the 300 K activated region with the 0 K electrodes. c) Electronic transmission coefficients $T_{300K}^{\parallel}(E,L)$ between the left and right electrodes calculated from ten MD snapshots (black lines) and their average $\langle T_{300K}^{\parallel}(E,L)\rangle$ (red line) used in the MD-Landauer approach (see main text). $E_c$ denotes the conduction band edge, and $E_F$ has been set 0.25 eV within the conduction band to calculate transport properties. In this example we have considered meta-NPG with $L = 30$ nm.

The electronic structure of each device, composed of the central MD region and the attached electrodes (Figure 1b), is captured via suitable TB models equipped with bond-length-dependent hopping parameters. The main results of our study are based on a nearest-neighbor TB Hamiltonian, as typically used for graphene, in the form

Eq. (1) $$\hat{H} = \sum_{\langle i,j \rangle} t_{ij} \hat{c}_i^\dagger \hat{c}_j,$$

where $\hat{c}_i^\dagger$ ($\hat{c}_j$) is the creation (annihilation) operator for $p_z$ orbitals at site $i$ ($j$) and the sum runs over first nearest neighbours. The bond-length dependent hopping parameter $t$ is expressed as

Eq. (2) $$t_{ij} = t_0 \left( \frac{a^2}{|\vec{R}_i - \vec{R}_j|^2} \right),$$

where $t_0$ is 2.5 eV, $a$ is 1.42 Å, and $\vec{R}_i$ ($\vec{R}_j$) is the position vector for atom $i$ ($j$). A similar TB parameterization was previously used to model quantum transport in thermally disordered carbon nanotubes.[41] We have also considered an alternative 2nd nearest-neighbour parametrization in the SI, previously used to model polycrystalline graphene,[42] that leads to the same qualitative results (see sections S1-S3 and Table S2-S3 in SI). The electronic band structure of all materials and their response to increasing bi-axial in-plane strains (from 1% to 5%) resulting from both TB descriptions are in agreement with density functional theory calculations (see Figure S1-4 in SI and Methods section for further details). The Hamiltonians are constructed and stored to file using SISL.

To evaluate transport characteristics, we apply the equilibrium GF formalism, as implemented in the open-source TBtrans code.[39] This method enables us to compute the transmission probability between the two attached electrodes, $T_{300K}(E, L)$, which is a spectral function (defined at each energy point, $E$) that depends on the atomic fluctuations in the device region at 300 K and its length $L$ (i.e., distance between electrodes; see Figure 1b). As depicted in Figure 1c, we average over ten MD snapshots to obtain the mean transmission function per material and device length, $\langle T_{300K}(E, L) \rangle$, from which we then extract the length-dependent conductance, $G_{300K}(L)$ from the Landauer formula,[35]

Eq. (3) $$G_{300K}(L,\mu) = G_0 \int \langle T_{300K}(E,L,\mu)\rangle \left(\frac{\partial f(E,\mu,300K)}{\partial E}\right) dE,$$

where $G_0 = \frac{2e^2}{h}$ is the quantum of conductance, $e$ is the elementary charge, $h$ is Planck's constant and $f(E,\mu,300K)$ is the Fermi-Dirac distribution at chemical potential $\mu$. We set $\mu$ to be 0.25 eV above the NPG conduction band edge $E_C$ (0.25 eV above the Dirac point for graphene), therefore focusing on electron transport. However, given the electron-hole symmetry in the band structure of all considered materials (see Figure S1-4 in SI), our main findings are equally applicable to hole transport. From the conductance we extract the device resistance which, in case of diffusive transport, varies linearly with length,

Eq. (4) $$R_{300K}(L,\mu) = R_c + \rho_{1D}(\mu) \cdot L = \frac{1}{G_{300K}(L,\mu)},$$

where $R_c$ is the contact resistance. From the slope of $R_{300K}(L,\mu)$, i.e., the one-dimensional resistivity $\rho_{1D}$ in units of [Ω/m], we extract the bulk resistivity in units of [Ω], $\rho_{bulk}(\mu) = \rho_{1D}(\mu) \cdot W$, with $W$ being the device width (see Table S1). This procedure is done for each in-plane direction, resulting in resistivity values for the direction parallel to the GNRs ($\rho_{bulk}^{\parallel}$) and perpendicular to them ($\rho_{bulk}^{\perp}$). The direction-dependent conductivity is then given by

Eq. (5) $$\sigma^{\parallel,\perp}(\mu) = \frac{1}{\rho_{bulk}^{\parallel,\perp}(\mu)},$$

which can in turn be used to estimate the mean free path *via*[43]

Eq. (6) $$l(\mu) = \frac{2\sigma(\mu)}{e^2 \cdot DOS(\mu) \cdot v_F(\mu)}.$$

In Eq. 6, $e$ is the elementary charge while $DOS(\mu)$ and $v_F(\mu) = \nabla_k E(k)/\hbar$ are the density of states and band velocity, respectively, evaluated at the chemical potential $\mu$, with $\hbar$ being the reduced Planck's constant.

### 3.2 Quantum transport along GNRs

We start by discussing transport along the y-direction, that is, parallel to the GNRs. Figure 2a-d shows the $\langle T_{300K}^{\parallel}(L)\rangle$ spectra with increasing device length for graphene (used as a reference) and the various NPGs. Each $\langle T_{300K}^{\parallel}(L)\rangle$ is obtained by averaging over ten MD samples (see Figure S5). We also show the pristine case, $T_{0K}^{\parallel}$, as dashed lines in Figure 2a-d. We see that thermal vibrations decrease the electronic transmission, and so $T_{0K}^{\parallel} > \langle T_{300K}^{\parallel}(L) \rangle$ for all cases. Likewise, as anticipated, increasing the device length (i.e., $L$ in Figure 1b) for the thermally activated samples (solid lines in Figure 2a-d) monotonically reduces $\langle T_{300K}^{\parallel}(L) \rangle$. All NPGs yield $\langle T_{300K}^{\parallel}(L) \rangle$ values which are of similar order of magnitude as in graphene. Also, like in graphene where electron-phonon coupling is low, $\langle T_{300K}^{\parallel}(L) \rangle$ in each NPG is within the same order of magnitude as $T_{0K}^{\parallel}$, which highlights that transport along the GNRs is not dramatically degraded due to vibrational disorder. This highlights the robustness of transport along the GNR channels against finite temperature.

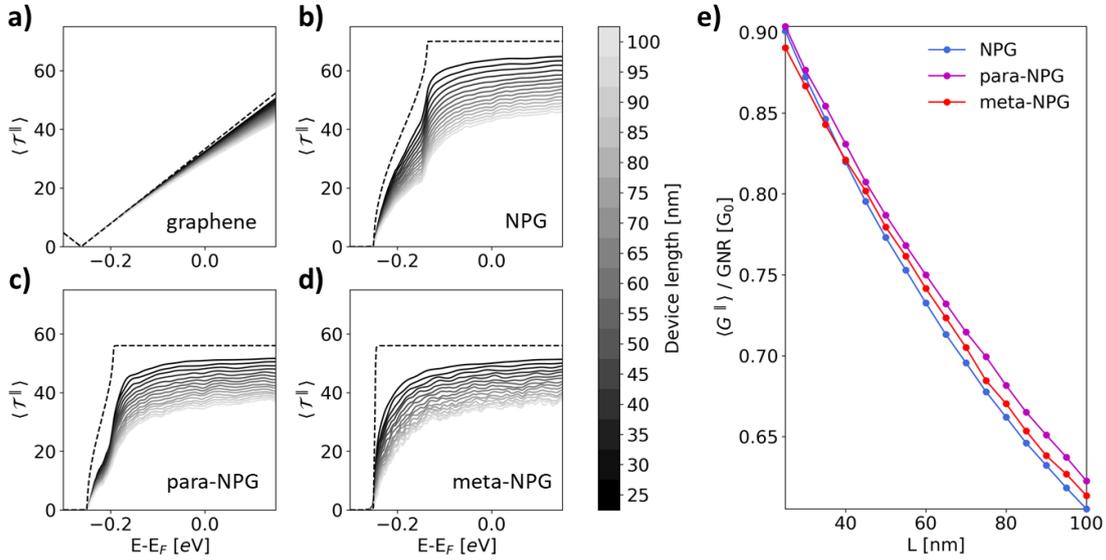

Figure 2. Averaged electronic transmission at 300 K, $\langle T_{300K}^{\parallel}(L) \rangle$, for a) graphene, b) NPG, c) para-NPG and d) meta-NPG for varying device lengths from 25 to 100 nm (solid lines). Different length values are represented with a grayscale. Dashed lines display $T_{0K}^{\parallel}$ – i.e. the transmission for the pristine systems at 0 K. e) Variation of $G_{300K}(L)$ with device length (L), at $\mu$ set at 0.25 eV above the conduction band edge, for the various considered materials, as calculated using Eq. (3) and normalizing by the number of GNRs included in the device geometries.

In Figure S6 in the SI we also see that all NPGs display approximately the same $T^{\parallel}$/GNR (normalized per GNR channel) both at 300 K and 0 K, with a similar dependence on device length. This suggests that the varying degree of inter-GNR electronic coupling in the different NPGs does not influence the effectiveness of quantum transport along GNRs at 300 K. Some differences between the NPGs may be noticed at the conduction band onset, both for $T_{0K}^{\parallel}$/GNR and $\langle T_{300K}^{\parallel}(L)\rangle$/GNR, which are due to the different inter-GNR couplings. However, the normalized transmission plateau at increasing device lengths (L) is almost exactly the same for all NPGs (Figure S6). Consequently, the $G_{300K}^{\parallel}(L)$/GNR curves for the three NPGs nearly overlap (see Figure 2e), resulting in very similar conductivity values, as shown in Table 1 (see Table S2 for the analogous results with an alternative TB parametrization).

Table 1. Conductivity at 300 K along the GNR direction for each material, calculated using Eq. (5).

|  | graphene | NPG | para-NPG | meta-NPG |
|---|---|---|---|---|
| $\sigma^{\parallel}$ (mS) | 13.90 | 6.85 | 6.17 | 6.11 |

### 3.3 Quantum transport across GNRs

Having found that inter-GNR coupling has minimal impact on transport along the GNRs, we now evaluate its effect on transport across the GNRs. As previously mentioned, all NPGs display very similar $T_{0K}^{\parallel}$ spectra with identical plateaus (per GNR), which ensures that approximately the same carrier density is being injected in the device region for the different NPG types. This, in turn, ensures that any divergence in the final σ∥ values between the considered NPG materials exclusively arises from the behaviour of each material at 300 K. This scenario, however, is not present for transport across GNRs. As shown in Figure 3a, the electronic transmission across GNRs at 0 K ($T_{0K}^{\perp}$) is very different between the three NPGs, with the meta-NPG displaying nearly zero $T_{0K}^{\perp}$ values for the entire considered energy range. This implies that using NPG, para-NPG and meta-NPG pristine electrodes within the NPG, para-NPG and meta-NPG devices, respectively, does not allow a direct comparison of their transport properties at 300 K, as the incoming electrode carrier density is massively different from case to case. To circumvent this issue, we utilize electrodes made of the same material for all three NPG devices, thus ensuring a fair comparison between materials at 300 K. We utilize NPG as our electrode material, as it yields the highest $T_{0K}^{\perp}$ (see Figure 3a). Since all NPGs are made of the same 7-13-AGNR, they have the same periodicity along the GNR direction (their unit cell length differs by less than 0.3%; see Table S1), which allows us to create smooth NPG/para-NPG and NPG/meta-NPG junctions, as displayed in Figure 3b.

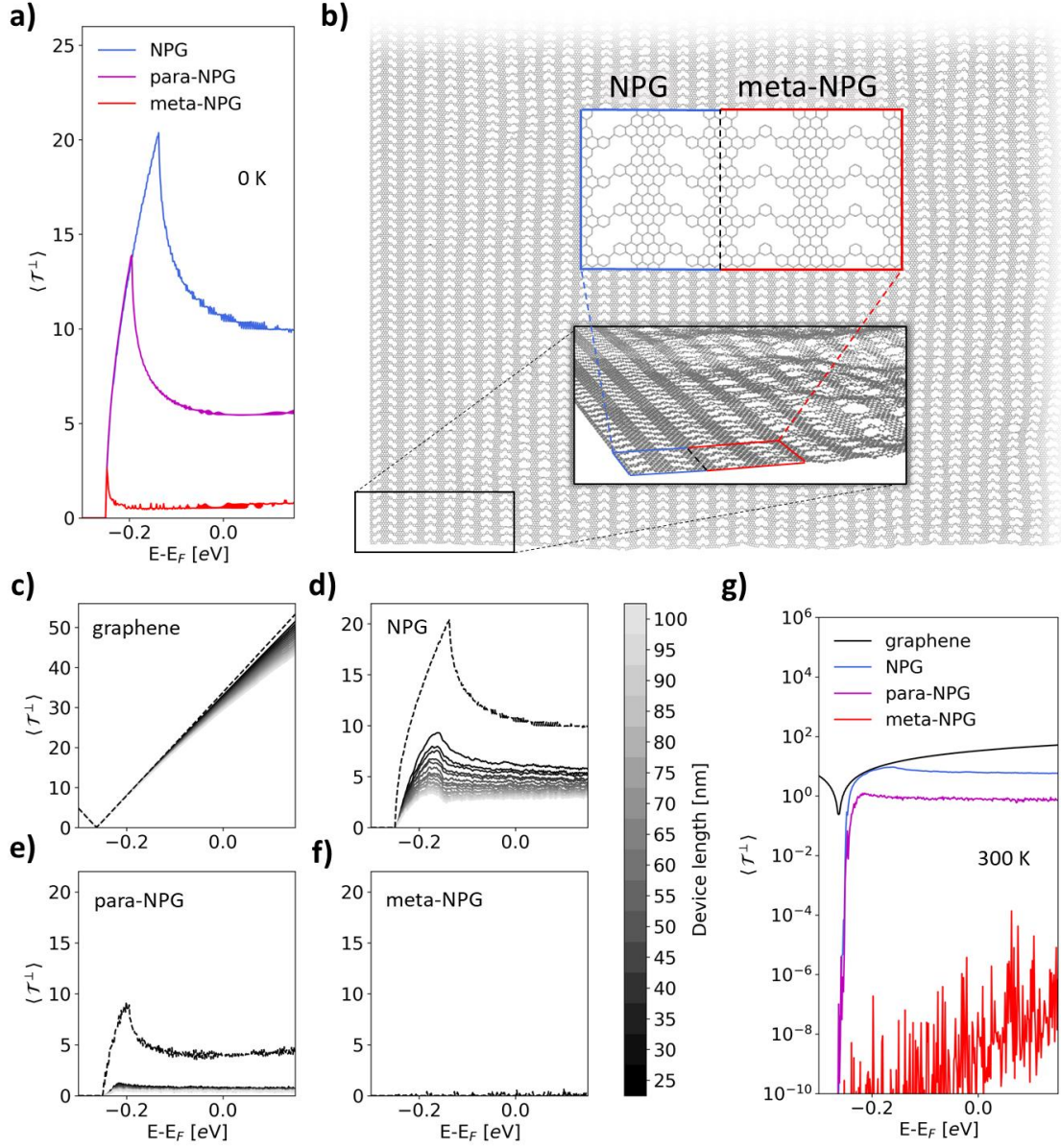

Figure 3. a) $T^{\perp}_{0K}$ for NPG, para-NPG and meta-NPG. b) Outline of a hybrid device where a 300 K meta-NPG sample is merged with a pristine NPG electrode. As shown in the inset, the NPG interface is at a GNR with distinct connections at either side. $\langle T^{\perp}_{300K}(L) \rangle$ for c) graphene, d) NPG, e) para-NPG and f) meta-NPG, for device lengths varying from 25 to 100 nm (solid lines). Dashed lines display $T^{\perp}_{0K}$ for each NPG. All spectra shown in panels c-f are computed using NPG leads. g) $\langle T^{\perp}_{300K} \rangle$ at L = 25 nm for the different materials, on a log scale. We note that transmission values below 10$^{-10}$ are outside of the precision range of our calculations, and so they may be considered as zero.

Figure 3c-f shows the averaged cross-GNR transmission at 300 K, $\langle T^{\perp}_{300K}(L) \rangle$, for increasing device lengths, *L*. As seen in Figure 3c, graphene displays the same behaviour

as for the other in-plane direction (Figure 2a), in full agreement with its isotropic 2D structure. On the contrary, because of their unique structure as 2D arrays of GNRs, all NPGs display significant transport anisotropy at 300 K, $T^{\parallel}_{300K} \gg T^{\perp}_{300K}$, as seen by comparing Figs. 2 and 3 (and as summarized in Figure S7 of the SI). This is in agreement with a previous study modelling electron-phonon coupling in the NPG.[44] Contrary to $T^{\parallel}$ (Figure 2), transport across GNRs shows notable differences between the different NPGs. The para-NPG displays the same qualitative behaviour as NPG, but with nearly half of its $T^{\perp}$ at 0 K and a significantly lower $T^{\perp}$ at 300 K (compare Figs. 3d and 3e). This demonstrates that, though para-connected phenyl rings are known to be conductive,[8] substitution of a C-C bond by a phenyl ring reduces current flow, especially under thermal fluctuations. More significantly, meta-NPG displays an even more extreme picture – namely that $T^{\perp}$ vanishes completely (Figure 3f), only displaying a negligible signal associated to $T^{\perp}_{0K}$. This demonstrates that QI embedded within meta-NPG, meant to cut transport across GNRs, is fully operational under thermal vibrations at 300 K. In fact, by plotting $\langle T^{\perp}_{300K}(25nm)\rangle$ on a logarithmic scale, as shown in Figure 3g, we see that it is only for the meta-NPG that transport across GNRs appears to be massively suppressed, with $\langle T^{\perp}_{300K}(25nm)\rangle \approx 10^{-8}$. Therefore, contrary to the other NPGs, meta-NPG effectively behaves as an electrical insulator in the direction perpendicular to the GNRs (Figure 3f-g) while behaving as a semiconductor in the other in-plane direction (Figure 2d-e). As shown in Figure S8, finite temperature seems to enhance the QI-induced GNR decoupling rather than deactivating it. This confirms the potential use of QI to engineer current flow in NPGs and, more generally, in carbon nanostructures at room temperature.

### 3.4 Overall picture

Finally, we take a closer look at the nature of transport in the studied materials, to provide further insight into their similarities and differences. First, we find that in all cases except for cross-GNR transport in meta-NPG, transport is in the diffusive, or ohmic, regime. This can be seen in Figure S9, where in all cases (except cross-GNR in meta-NPG) the resistance increases linearly with length. Following Eqs. (4)-(6), we report in Table 2 the mean free path in all materials for transport parallel and perpendicular to the GNRs. Here we see that transport along GNRs is very good, with mean free paths of several hundred nanometers, indicating the weak effect that thermal disorder has on transport in this direction. We also note that transport along the GNRs is barely affected by inter-ribbon coupling (see also Table 1 and filled bars in Figure 4) and hence thermal disorder does not seem to induce back scattering. This is different from what occurs with local electrostatic disorder, which leads to significantly degraded transport along GNRs upon approaching the "1D-like" transport of meta-NPG.[28] Meanwhile, the mean free path is dramatically suppressed for cross-GNR transport, and we see that the inter-ribbon coupling plays a significant role, with $l^\perp$ four times smaller in para-NPG compared to NPG.

Table 2. Mean free path at 300 K along the two in-plane directions for each material, calculated using Eq. 6.

|  | Graphene | NPG | para-NPG | meta-NPG |
|---|---|---|---|---|
| $l^\parallel$ (nm) | 772.6 | 277.8 | 317.4 | 318.4 |
| $l^\perp$ (nm) | 766.5 | 11.45 | 2.85 | 0* |

* Note: the mean free path was extracted from the resistivity at a length of 25 nm, and is equal to $8.5 \cdot 10^{-7}$ nm, which may be considered 0 within the accuracy of our model.

In contrast to all other cases, cross-GNR transport in meta-NPG is found to be strongly localized, with the resistance increasing exponentially with length. In this regime we cannot extract a mean free path because the resistivity, $\rho_{bulk}(\mu) = R_{300K}(L,\mu) \cdot W/L$, is not constant. But applying this expression to a device length of 25 nm gives us a mean free path of $8.5 \times 10^{-7}$ nm, which is effectively equal to 0. Meanwhile, fitting the resistance to an exponential, $R_{300K}(L,\mu) = R_c \cdot exp(L/\xi)$, yields a localization length of $\xi = 6.5$ nm, highlighting the extremely strong suppression of cross-GNR transport due to the meta-connected bridges inducing QI.

In Figure 4 we provide an overview of our results by comparing the electrical resistivities along ($\rho^\parallel$) and across GNRs ($\rho^\perp$). The $\rho^\parallel$ is very similar between all NPGs and is, in turn, of the same order as graphene's, indicating that GNRs within NPGs act as good conductors in all cases. On the other hand, $\rho^\perp$ increases by several orders of magnitude along the NPG series, which permits one to disentangle the effect of the different relevant structural parameters coming into play, namely: i) the anisotropic structure of NPGs, ii) the separation of GNRs with phenyl bridges and, most of all, iii) the QI.

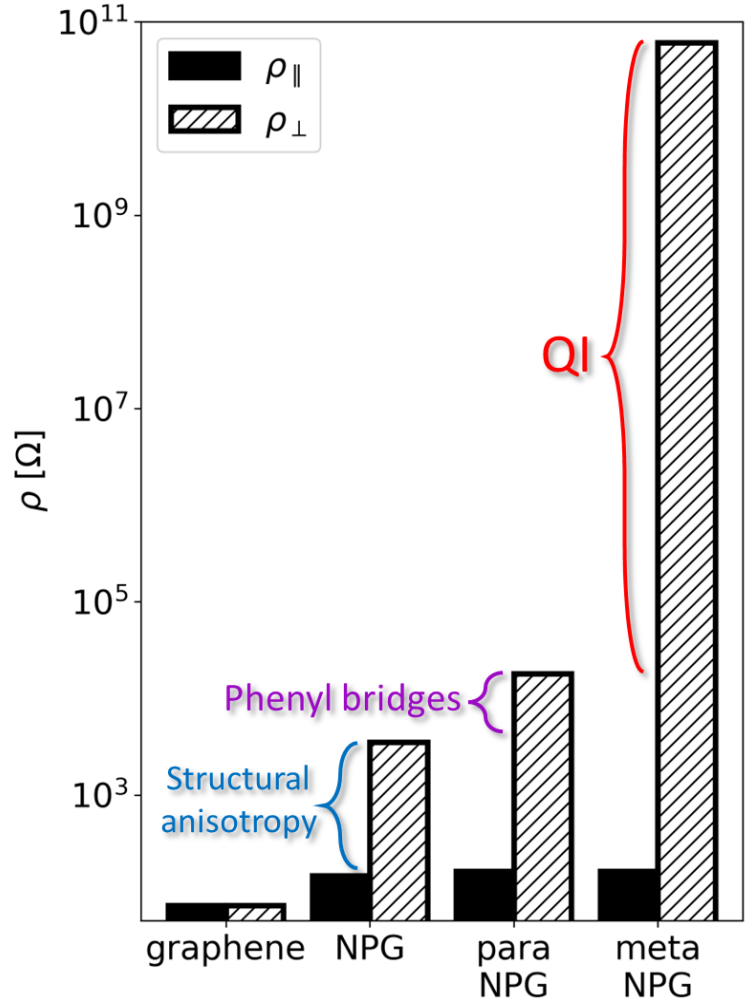

Figure 4. Electrical resistivities along GNRs ($\rho^{\parallel}$, filled bars) and across GNRs ($\rho^{\perp}$, hatched bars) for each considered material at 300 K. The different parameters affecting $\rho^{\perp}$ are indicated in curly brackets. The cross-GNR resistivity of meta-NPG was extracted at a length of 25 nm.

**4. Conclusions**

In this work, by combining MD simulations and quantum transport calculations we have shown that QI, long studied in single-molecule electronics, may be used to tailor current flow in carbon nanostructures such as NPGs under thermal fluctuations at 300 K. To do this we have characterized the transport properties of various NPGs, featuring different inter-ribbon connections, along the in-plane directions parallel and perpendicular to the GNRs. We have used TB models with bond-length-dependent hopping terms, thus capturing the effect of thermal disorder in the electronic properties of each system. The resulting large-scale NPG samples, fed with such TB Hamiltonians, have been coupled to pristine semi-infinite electrodes, allowing the computation of electronic transmission and conductivity using the GF formalism. We have found that in the direction parallel to GNRs all NPGs are good conductors at 300 K, exhibiting diffusive transport and mean free paths on the order of hundreds of nm regardless of their inter-ribbon connection. On the contrary, transport across GNRs shows conductivities well below graphene's and is entirely determined by the particular GNR connection. While the anisotropic structure of NPGs is the first cause of transport anisotropy for all NPGs, substitution of the C-C bond in NPG by a para-connected ring, as in para-NPG, further reduces inter-ribbon coupling, and so transmission, especially under the effect of thermal vibrations. This result implies that heavily suppressed transport should be expected across GNRs featuring bridges composed of two phenyl rings or more, as in recently fabricated phenylated NPGs.[27] However, it is in the meta-NPG where transport across GNRs completely vanishes, reaching resistivity ($\rho^\perp$) values above $10^{10}$ Ω along that in-plane direction. This result, unique to meta-NPG, is a direct manifestation of the persistence of QI in the meta-configured phenyl bridges embedded in this nanomaterial. This, in turn, unequivocally demonstrates that QI-engineering may be used at room temperature to tailor current flow within carbon nanostructured materials. Our results therefore highlight the fundamental role that QI may play in the future to realize carbon nanocircuitry and, more generally, carbon nanoelectronics.

## 5. Code and data availability

All the input files and data reported in this article are freely available on Zenodo (https://doi.org/10.5281/zenodo.11443216).

**Conflicts of interest**

No conflicts of interest to declare.

**Acknowledgements**

I.A. is grateful for a Juan de la Cierva postdoctoral grant (FJC2019-038971-I) from the Ministerio de Ciencia e Innovación (MCIN). G.C. acknowledges the project SAMOTHRACE "Sicilian micro and nanotechnology research and innovation centre" founded by PNRR-MUR (ECS_00000022, CUP B63C22000620005) for partial support.

O.K. is supported by the REDI Program, a project that has received funding from the European Union's Horizon 2020 research and innovation program under the Marie Sklodowska-Curie grant agreement number 101034328. We thank the DTU Computing Center for HPC resources.[45] A.C. and S.R. acknowledge support from PID2019-106684GB-I00 also funded by MCIN/AEI/10.13039/501100011033/FEDER, UE, as well as PID2022-138283NB-I00 funded by MICIU/AEI/10.13039/501100011033 and SGR funded by Generalitat de Catalunya. ICN2 is funded by the CERCA Programme from Generalitat de Catalunya, and is currently supported by the Severo Ochoa Centres of Excellence programme, Grant CEX2021-001214-S, both funded by MCIN/AEI/10.13039.501100011033.